\documentclass[aip,prb,showpacs,twocolumn]{revtex4}

\usepackage{graphics,graphicx,amssymb,amsmath,color}
\begin{document}

\title{Thermodynamic Properties of the Quantum Spin Liquid Candidate ZnCu$_{3}$(OH)$_{6}$Cl$_{2}$ in High Magnetic Fields}
\author{Tian-Heng~Han$^{1,2,3\ddag}$}
\author{Robin~Chisnell$^{1}$}
\author{Craig J.~Bonnoit$^{1}$}
\author{Danna E.~Freedman$^{4,5}$}
\author{Vivien S.~Zapf$^{6}$}
\author{Neil~Harrison$^{6}$}
\author{Daniel G.~Nocera$^{4}$}
\author{Yasu~Takano$^{7}$}
\author{Young S.~Lee$^{1,\dagger}$}
\affiliation{$^{1}$Department of Physics, Massachusetts Institute of Technology, Cambridge, Massachusetts 02139, USA}
\affiliation{$^{2}$The James Franck Institute and Department of Physics, The University of Chicago, Chicago, Illinois 60637, USA}
\affiliation{$^{3}$Materials Science Division, Argonne National Laboratory, Argonne, Illinois 60439, USA}
\affiliation{$^{4}$Department of Chemistry, Massachusetts Institute of Technology, Cambridge, Massachusetts 02139, USA}
\affiliation{$^{5}$Department of Chemistry, Northwestern University, Evanston, Illinois 60208, USA}
\affiliation{$^{6}$National High Magnetic Field Laboratory, Los Alamos National Laboratory, Los Alamos, New Mexico 87545, USA}
\affiliation{$^{7}$Department of Physics, University of Florida, Gainesville, Florida 32611, USA}
\date{\today}

\begin{abstract}

We report measurements of the specific heat and magnetization of single crystal samples of the spin-1/2 kagome compound ZnCu$_{3}$(OH)$_{6}$Cl$_{2}$ (herbertsmithite), a promising quantum spin-liquid candidate, in high magnetic fields and at low temperatures. The magnetization was measured up to $\mu_{0}H$ = 55 T at $T$ = 0.4 K, showing a saturation of the weakly interacting impurity moments in fields above $\sim10$~T. The specific heat was measured down to $T < 0.4$~K in magnetic fields up to 18 T, revealing $T$-linear and $T$-squared contributions. The $T$-linear contribution is surprisingly large and indicates the presence of gapless excitations in large applied fields. These results further highlight the unusual excitation spectrum of the spin liquid ground state of herbertsmithite.\end{abstract}

\pacs{75.10.Kt  75.30.Gw  75.40.Cx  75.50.Ee} \maketitle

Quantum spin liquids represent a fundamentally new state of matter whose ground state is not characterized by a local order parameter. However, finding them as the ground state of real materials has been a great challenge. The physics of quantum spin liquids may be of relevance in the understanding of high T$_{c}$ superconductivity\cite{Anderson1987, Lee2006} as well as applications in quantum computation\cite{Ioffe}. The $S$=$\frac{1}{2}$ Heisenberg antiferromagnet on the kagome lattice (composed of corner sharing triangles) has long been an ideal system in which to look for spin-liquid physics due to the high degree of frustration and low value of the spin\cite{Sachdev, MisguichLhuillier, Elser1989}. Recent numerical calculations based on the nearest neighbor Heisenberg spin Hamiltonian point to a fully gapped spin-liquid ground state\cite{White,Depenbrock}. For the most promising candidate materials, it is difficult to prove the presence of a spin-liquid ground state since current experimental techniques do not directly couple to the topological order which characterizes the state. However, quantum spin liquids supports exotic quantum fractionized excitations such as spinons\cite{Balents}. Hence, probing the low energy spin excitations and comparing with theoretical predictions can serve as an effective method to uncover new physics.

The $x=1$ end member of the family Zn paratacamite [Zn$_{x}$Cu$_{4-x}$(OH)$_{6}$Cl$_{2}$], called herbertsmithite, is one of the most promising materials to have a quantum spin liquid ground state\cite{Shores}. This system consists of kagome planes of Cu$^{2+}$ ions separated by layers of nonmagnetic Zn$^{2+}$ ions. While the Cu-O-Cu antiferromagnetic superexhange interaction is estimated to be $J\sim$17~meV, no magnetic transition or long-range ordering has been observed down to $T=50$~mK\cite{Helton,Mendels,Imai}. Our recent success in the growth of large high quality single crystal samples has greatly advanced the knowledge of herbertsmithite as a spin liquid material\cite{Han, Han2, Freedman, Wulferding, Ofer3, Imai2, Pilon}. In fact, the direct observation of a spinon continuum using inelastic neutron scattering on a single crystal sample has been achieved, which is a signature of the quantum spin liquid\cite{Han3}. The underlying spin Hamiltonian is believed to be predominantly Heisenberg exchange with DM and easy-axis anisotropy as additional perturbations\cite{Zorko, Han2}. Single crystal anomalous x-ray refinement confirms the absence of non-magnetic dilution in the kagome layers\cite{Freedman}, although an excess of $\sim$5\% Cu$^{2+}$ ions resides on the interlayer sites. These weakly coupled spin impurities produce a magnetic response which become significant at low energy and temperature scales, masking the intrinsic behavior.

For probing spin liquids, specific heat measurements are useful in that they are sensitive to the total density of states. In case where a triplet spin-gap is believed to exist, a large number of lower lying singlet states may fill the gap\cite{Waldtmann}. The density of states of low energy excitations should provide crucial clues in elucidating the spin liquid ground state. In prior specific heat measurements at low temperatures\cite{Helton}, powder samples of herbertsmithite were mixed with grease to improve the thermal contact with the probe. By using single crystal samples, one can reduce the volume of grease used and, hence, improve the measurement accuracy. This is especially important when performing measurements in strong applied magnetic fields where the magnetic contribution to the specific heat at low temperatures is strongly reduced, as we do here. By suppressing the impurity contribution to the specific heat, our results reveal that the intrinsic behavior of $C(T)$ has a significant $T$-linear term, which is surprising for an insulating magnet. In addition, a $T$-squared contribution is also observed. The combined results underscore the exotic spin liquid state that describes herbertsmithite.

\begin{figure}
\centering
\includegraphics[width=15.7cm]{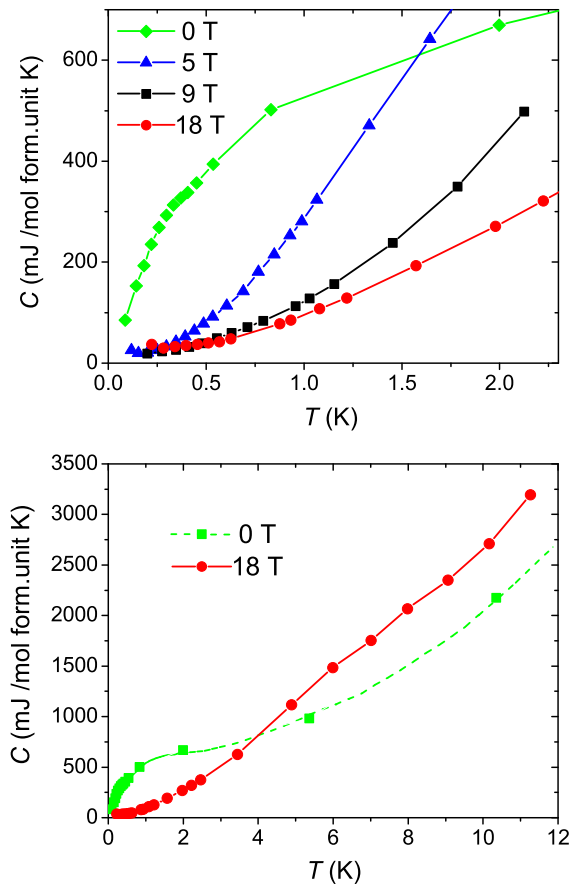} \vspace{-4mm}
\caption{(color online) (a) Specific heat of a single crystal sample of herbertsmithite measured in various applied magnetic fields parallel to the kagome plane. (b) Comparison of the specific heat at 0 T and 18 T over an extended temperature range. Based on measurements at zero field on a protonated sample, the 0 T data is supplemented by the green dashed line.} \vspace{-4mm}
\label{Figure1}
\end{figure}

We have performed specific heat measurements using a standard relaxation method on a high quality deuterated single crystal sample of herbertsmithite down to dilution fridge temperatures. The experiments were performed at the National High Magnetic Field Laboratory (NHMFL) at Tallahassee using the 18 T superconducting magnet SCM1. The sample was grown and characterized as reported previously\cite{Han, Han3}. In order to optimize thermal contact, the crystal was shaped into a thin slab with dimensions 1.5 x 1 x 0.4 mm$^{3}$ (weighing 2.2 mg). A minimum amount of grease, $\sim$ 0.2 mg, has been used to attach the sample. The applied magnetic field was oriented within the kagome plane ($H$$\perp$$c$). The specific heat data in various applied fields are plotted as a function of temperature in Figure 1(a). Magnetic fields of 5 T and higher greatly suppress the low temperature specific heat, pushing the change in entropy to higher temperatures. The suppressed specific heat likely arises from the excess impurity spins on the interlayer sites. The vanishing difference between the 9 T and 18 T data at $T$ $\leq$ 1 K indicates an effective saturation of impurity moments. Hence, the specific heat at high fields below 2 K is predominantly due to the intrinsic spins on the kagome layers.  In Figure 1(b), the data at 0 T and 18 T are plotted up to $T$ $\sim$ 10 K. The field-driven upshift in temperature of $C(T)$ is clearly evidenced. No indication of a phase transition exists for 0.2 $\leq$ $T$ $\leq$ 11 K at $\mu_{0}H$ = 18 T.

\begin{figure}
\centering
\includegraphics[width=7.5cm]{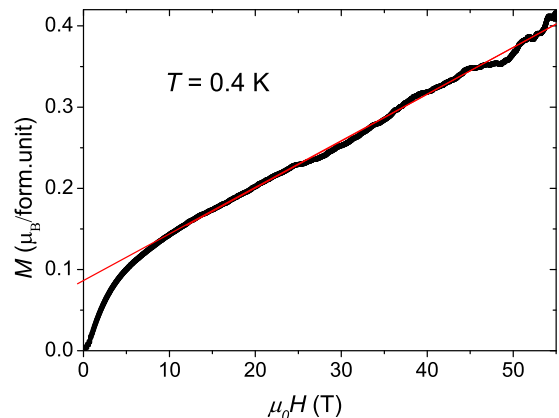} \vspace{-4mm}
\caption{(color online) Magnetization versus magnetic field measured on a single crystal sample of herbertsmithite. The field is parallel to the kagome plane. A straight line is fitted for $\mu_{0}H$ $>$ 10 T.} \vspace{-4mm}
\label{Figure2}
\end{figure}

In Figure 2, the magnetization of a single crystal sample is plotted as a function of field up to 55 T. The measurements were performed at the NHMFL at Los Alamos using the 65 T pulsed magnet with a pulse duration of $\sim$25 ms at $T = 0.4$ K. Two protonated single crystal samples of herbertsmithite were coaligned with a total mass of 1.6 mg. For large fields, $\mu_{0}$$H$ $>$ 10 T, the magnetization $M$ is linear in field $H$. This indicates that the impurity moments become saturated above this field strength (at $T = 0.4$ K). The magnetization value per formula unit was normalized by comparing d$M$/d$H$ at high fields with previous measurements on similar samples\cite{Han, Helton2}. The value of d$M$/d$H$ was measured at $T$ =1.9 K and $\mu$$_{0}H$ = 14 T using a similar crystal, and the AC susceptibility of a powder sample was used to determine the temperature dependence in order to extrapolate to $T$ = 0.4 K for the normalization. The vibration of the instrument with increasing pulsed magnetic fields leads to a reduction in the signal to noise ratio at the highest measured fields. However below 25 T, no field-induced phase transition is observed. In a strong magnetic field, the interlayer impurity spins can be modeled by doublets with $g \simeq 2.2$ and a zero-field splitting of $\sim$ 1 K\cite{Bert, deVries}. The Schottky anomaly due to these impurities becomes negligible when $k_{B}T$/$\mu$$_{B}H$ $<$ 0.1, meaning a 18 T field effectively suppresses the impurity specific heats below 2 K. The phonon contribution at these low temperatures is negligibly small. If we assume the specific heat at temperatures higher than $T = 30$ K\cite{Helton} is dominated by phonons, then the $T^{3}$ temperature dependence indicates that at $T$ = 1 K it is less than 1 \% of the total specific heat.

In Figure 3, $C/T$ is plotted versus temperature for three applied fields. The upturn that is seen with decreasing temperature (at very low temperatures) is due to the Schottky effect from the nuclear moments. In panel (a), the data are fitted to $C$/$T$ = $\gamma$ + $\alpha$T (the lower bound of the fitted temperature range is indicated by the arrow) and the results are listed in Table 1. The low temperature specific heat is very well described by $T$-linear and $T$-squared terms. We note that the fitted values for $\gamma$ are substantial and appear to saturate at the highest fields (between 9 T and 18 T). In Figure 3(b), additional fits were performed by explicitly including the Schottky anomaly from nuclear magnetic dipole moments. The temperature scales of nuclear dipole splittings in a 18 T field are much smaller than our measurement range. Thus, their contribution to $C$/$T$ can be well approximated as $AH^{2}$/$T^{3}$. The data are fitted to $C$/$T$ = $AH$$^{2}$/$T^{3}$ + $\gamma$ + $\alpha$$T$ and the results, also listed in Table 1, are consistent with the previous fits. The values of $A$ at 18 T and at 9 T are nearly identical, consistent with the nuclear Schottky origin of the low temperature upturn in $C$/$T$. No evidence of an exponentially activated gap-like behavior is seen in the specific heat down to $T$ $\sim$ 0.5 K ($\sim$ $J$/400).

\begin{figure}
\centering
\includegraphics[width=11.8cm]{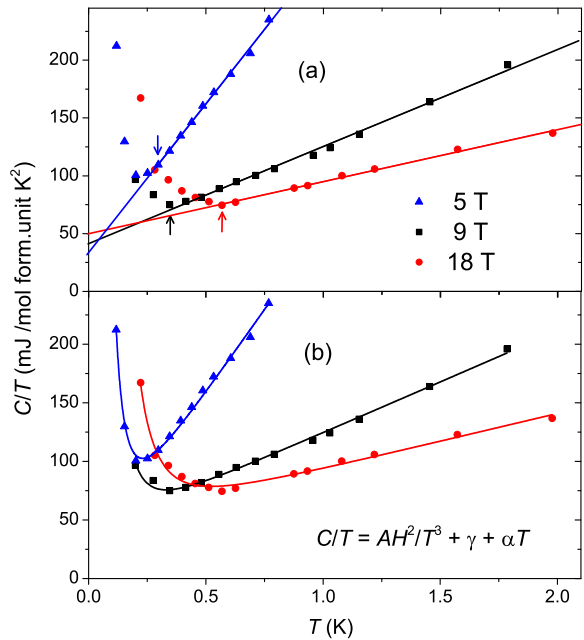} \vspace{-4mm}
\caption{(color online) The specific heat divided by temperature measured on a deuterated single crystal of herbertsmithite. (a) Linear fits with a $T$-linear and a $T$-squared terms. The arrows indicate the lowest temperatures included in the fittings. (b) Polynomial fits with a nuclear Schottky, a $T$-linear and a $T$-squared terms.} \vspace{-4mm}
\label{Figure3}
\end{figure}

\begin{table}
\caption{Returns from the fits of $C/T$ in Figure 3 measured on the deuterated crystal of herbertsmithite. The results are normalized to per mole formula unit, where each formula unit contains three Cu$^{2+}$ spins.}

\vspace{4mm}

\begin{tabular}{| c | c | c | c |}
\hline
$C$/$T$ = $\gamma$ + $\alpha$$T$ & 5 T& 9 T & 18 T\\
\hline
$\gamma$  (mJ/mol K$^{2}$) &33(3) & 42(2) & 50(1)\\
$\alpha$  (mJ/mol K$^{3}$) &258(6) & 84(2) & 45(1)\\
\hline
adjusted R$^{2}$ &0.9959 & 0.9938 & 0.9948\\
\hline
\end{tabular}

\vspace{4mm}

\begin{tabular}{| c | c | c | c |}
\hline
$C$/$T$ = $AH$$^{2}$/$T^{3}$ + $\gamma$ + $\alpha$$T$ & 5 T& 9 T & 18 T\\
\hline
$A$  ($\mu$J K/mol T$^{2}$) & 10.7(3) & 4.3(3) & 3.7(1) \\
$\gamma$  (mJ/mol K$^{2}$) &15(3) & 37(2) & 45(3)\\
$\alpha$  (mJ/mol K$^{3}$)& 285(7) & 87(2) & 48(2)\\
\hline
adjusted R$^{2}$ &0.9933 & 0.9945 & 0.9807\\
\hline
\end{tabular}

\label{Table1}
\end{table}

Specific heat data measured with $H$$\perp$$c$ and $H$$\parallel$$c$ are plotted in Figure 4. The measurements were performed on a 4.1 mg protonated single crystal sample\cite{Han2} of herbertsmithite using a Quantum Design PPMS. The magnetic susceptibility indicates that the amount of impurities is similar to that of the deuterated crystal. The specific heat data have been fitted to $C$/$T$ = $AH^{2}$/$T^{3}$ + $\gamma$ + $\alpha$$T$ and the resulting parameters are listed in Table 2. These values are reproducible between measurements on several other protonated crystals which were picked from different growth batches (not shown). No intrinsic magnetocaloric anisotropy is observed, consistent with the Heisenberg model being a good approximation.

For the combined data on our crystals (Tables 1 and 2), a basic observation that can be made is that the applied magnetic field enhances the $\gamma$ value and suppresses the $\alpha$ value. We note that there are some subtle differences between the protonated and deuterated samples. In Table 2, the $\gamma$'s at 14 T for both field orientations are larger than the ones in Table 1. At 9 T, discrepancies between protonation and deuteration appear in both $\gamma$ and $\alpha$. It is not clear whether these difference reflect differences in the underlying spin-liquid physics. The $A$ coefficient, in the nuclear Schottky term, is calculated to be $A_{H}$=62.42 $\mu$J K/mol f.u. T$^{2}$ for a protonated crystal and $A_{D}$=13.62 $\mu$J K/mol f.u. T$^{2}$ for a deuterated crystal. The largest contribution to $A_{H}$ comes from hydrogen. In Tables 1 and 2, the $A$ values are fractions of the calculated values, indicating that all nuclear spins have not fully relaxed. A strong nuclear Schottky magnetocaloric anisotropy, $A_{H\perp c}$ $<$ $A_{H||c}$, is apparent at 14 T. This likely indicates a longer $T_{1}$ for hydrogen (on the order of 10 seconds which is the relaxation time for the specific heat measurements) when the field is parallel to the kagome plane. The $A$ values are larger in Table 2 than in Table 1 since $A_{H} >$ $A_{D}$.

A $T$-linear specific heat is unusual in a 2D Mott insulator with a disordered magnetic ground state. Such a temperature dependence appears in spin glasses and Fermi-liquid excitations in metals; however, neither of these describe the spin system in herbertsmithite. The large linear coefficient, $\gamma$ = 45(3) mJ/mol f.u. K$^{2}$ for a deuterated crystal at 18 T and $\gamma$ = 72(3) mJ/mol f.u. K$^{2}$ for a protonated crystal at 14 T, indicate the presence of gapless excitations. One possibility is the presence of a spinon Fermi surface. An U(1) Dirac spin liquid should have a $T$-linear specific heat when $k_{B}T$ $<< \mu_{B}H$.\cite{Ran} The Fermi point on a Dirac node of spinons expands into a Fermi pocket in an applied field, providing a finite density of gapless excitations. This theory also predicts that $\gamma$ should increase with applied field which is consistent with our data. However, the specific prediction of $\gamma \propto$ $H$ (as the radius of the Fermi pocket is proportional to $H$) does not quantitatively match our data.

A $T$-squared temperature dependence is also not expected. While spin wave excitations on a 2D antiferromagnet give a $T^{2}$ specific heat, this is at odds with the spin disordered ground state in herbertsmithite. Previous numerical calculations on the spin-1/2 kagome Heisenberg model indicate a spin liquid ground state with an exponentially large number of singlet excitations filling the spin-gap\cite{Waldtmann}. These non-magnetic states may produce a field-independent $T^{2}$ specific heat\cite{Sindzingre}. An U(1) Dirac spin liquid\cite{Ran} (in zero field) and an algebraic vortex spin liquid\cite{Ryu} also yield a $T^{2}$ temperature dependence to the specific heat.

\begin{figure}
\centering
\includegraphics[width=7.5cm]{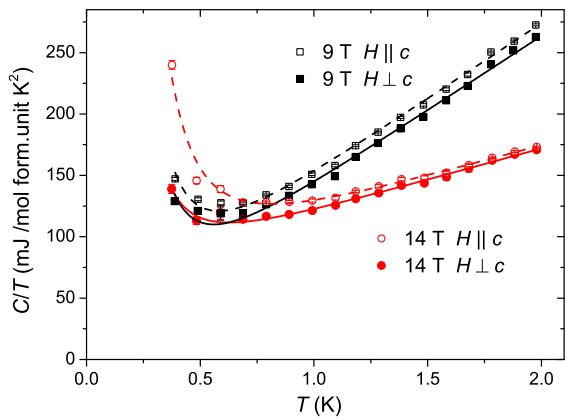} \vspace{-4mm}
\caption{(color online) Specific heat of a protonated single crystal of herbertsmithite measured with $H$$\perp$$c$ and $H$$\parallel$$c$, where the $c$-axis is normal to the kagome plane. The lines denote fits as described in the text.} \vspace{-4mm}
\label{Figure4}
\end{figure}

\begin{table}
\caption{Returns from the fits of $C/T$ in Figure 4 measured on a protonated crystal of herbertsmithite. The results are normalized to per mole formula unit, where each formula unit contains three Cu$^{2+}$ spins.}

\vspace{4mm}

\begin{tabular}{| c | c | c |}
\hline
$H$ $\perp$ $c$ & 9 T & 14 T\\
\hline
$A$  ($\mu$J K/mol T$^{2}$)& 49(4) & 14(1) \\
$\gamma$  (mJ/mol K$^{2}$) & 19(4) & 68(2)\\
$\alpha$  (mJ/mol K$^{3}$) &122(3) & 52(1)\\
\hline
adjusted R$^{2}$ & 0.9937 & 0.9938\\
\hline
\end{tabular}

\vspace{4mm}

\begin{tabular}{| c | c | c |}
\hline
$H$ $\parallel$ $c$ & 9 T & 14 T\\
\hline
$A$  ($\mu$J K/mol T$^{2}$) & 58(4) & 38(2) \\
$\gamma$  (mJ/mol K$^{2}$) &  25(3) & 72(3)\\
$\alpha$  (mJ/mol K$^{3}$)&  124(2) & 51(2)\\
\hline
adjusted R$^{2}$ & 0.9962 & 0.9857\\
\hline
\end{tabular}

\label{Table2}
\end{table}

These observations for the magnetic specific heat of herbertsmithite provide a point of comparison with the organic spin liquid materials. A $T$-linear dependence of the specific heat has been reported previously in $\kappa$-(BEDT-TTF)$_{2}$Cu$_{2}$(CN)$_{3}$\cite{Yamashita1} and EtMe$_{3}$Sb[Pd(dmit)$_{2}$]$_{2}$\cite{Yamashita2}. It is believed that this behavior may be related to excitations about a spinon Fermi surface. In fact, these specific heat observations in the organic materials were some of the strongest early evidence of spin liquid ground states. Interestingly, the large value of $\gamma$ ($\sim$ 50 mJ/mol f.u. K$^{2}$) that we measure in our samples of herbertsmithite are comparable to the values measured in the organics. However, our measurements $C(T)$ clearly show a field-dependence, in contrast to the field-independent behavior seen in the organics. Moreover, the theoretically expected ground states are likely different between the trianglar lattice and kagome lattice systems.  Previously, field-independent $T$-squared specific heat has been observed in the kagome-like compound SrCr$_{9p}$Ga$_{12-9p}$O$_{19}$\cite{Ramirez2}. And Cu$_{3}$V$_{2}$O$_{7}$(OH)$_{2}\cdot$2H$_{2}$O (vorborthite), a distorted spin-1/2 kagome lattice, has both field-dependent $T$-linear and $T$-squared magnetic specific heat contributions appearing in the same low temperature range\cite{Yamashitajpcm}. However, a phase transition exists at $T$ = 1 K below which magnetic order appears\cite{YoshidaH, YoshidaM}. Here, herbertsmithite has a spin liquid ground state in which both terms appear in the same low temperature range in high magnetic fields.

It is important to be mindful of several factors when interpreting the data. First, the reported values for $\gamma$ and $\alpha$ result from measurements with an applied field. It is not clear that these apply to the zero-field specific heat of the intrinsic kagome spins. The large applied fields are chosen to suppress the impurity contribution, however these fields may also be large enough to close any predicted spin-gap. If the spinons are fermions, the results may indicate that a spinon Fermi surface forms when a spin-gap closes at high fields. Second, while the gross impurity behavior may be suppressed with $\mu_{0}H$ $\geq$ 10 T at low temperatures, there could be subtle interactions involving impurities which remain (such as the coupling between the impurities and the in-plane spins). Third, subtle difference between protonated and deuterated crystals exist. It is possible that these are due to subtle differences in the crystal structure which produce subtle changes in the magnetic excitations.

Further experiments using single crystal samples of herbertsmithite should help shed light on some of the above questions. For example, the nature of the excitations in herbertsmithite can be further clarified by measuring its thermal conductivity which is sensitive to itinerant excitations. The field-dependence of the thermal conductivity and the presence/absence of a thermal Hall effect may help to distinguish between spinons\cite{Balents}, spinless vortices\cite{Ryu}, Majorana fermions\cite{Biswas} and other scenarios. Improved understanding of the impurity interactions would also help in our understanding of the low energy excitations. In conclusion, the magnetic specific heat of single crystal samples of herbertsmithite show both $T$-linear and $T$-squared terms at low temperatures and high fields. These results, together with the spinon continuum observed using neutron scatterings\cite{Han3}, make a compelling case for the quantum spin liquid state in herbertsmithite.

We thank Tim Murphy, Ju-Hyun Park and Glover Jones for assistance in the experiment, Shaoyan Chu for assistance in the sample preparation and Patrick Lee, Todadri Senthil, Michael Norman, Leon Balents, Subir Sachdev, Ming-Xuan Fu, Takashi Imai, Tyrel McQueen and Joel Helton for useful discussions.  This work was primarily supported by the Department of Energy (DOE) under Grant No. DE-FG02-07ER46134 (sample synthesis and thermodynamic measurements). We also acknowledge support from the NSF ACC-F under CHE-1041863. The NHMFL is supported by NSF Cooperative Agreement No. DMR-0654118 and DMR-1157490, by the State of Florida, and by the DOE. YT acknowledges support by the NHMFL UCQP. \\
$^{\ddag}$tianheng@alum.mit.edu\\
$^{\dagger}$younglee@mit.edu
\bibliography{LowTC}
\end{document}